\documentstyle[twocolumn,aps,pra,epsfig]{revtex}
\begin{document}
\wideabs{
\title{Critical velocity in cylindrical Bose-Einstein condensates}
\author{P.O. Fedichev\protect\( ^{1,3}\protect \) and G.V. Shlyapnikov\protect\( ^{2,3}\protect \)}
\address{\protect\( ^{1}\protect \)Institut f. Theoretische Physik,
Universit\"at Innsbruck, Technikerstr. 25/2, A6020, Innsbruck, Austria}
\address{\protect\( ^{2}\protect \)FOM Institute for Atomic and Molecular Physics,
Kruislaan 407, 1098SJ Amsterdam, The Netherlands}
\address{\protect\( ^{3}\protect \)Russian Research Center Kurchatov Institute, Kurchatov
Square, 123182 Moscow, Russia}
\maketitle
\begin{abstract}
We describe a dramatic decrease of the critical velocity in
elongated cylindrical Bose-Einstein condensates which originates from the
non-uniform character of the radial density profile. We discuss this 
mechanism with respect to recent measurements at MIT.
\end{abstract}
 }

Superfluidity is one of the striking manifestations of quantum statistics, and 
the occurence of this phenomenon depends on the excitation spectrum of
the quantum liquid. The key physical quantity is the critical velocity \(
v_{c} \), that is, the maximum velocity at which  the flow of the liquid is
still non-dissipative (superfluid). The well-known Landau criterion
\cite{LL:volIX} gives the critical velocity as the minimum ratio of energy to
momentum in the excitation spectrum: 
\begin{equation}  \label{Landaucrit}
v_{c}=\min \left( \frac{\epsilon (k)}{k}\right) 
\end{equation}
(we put \( \hbar=1 \) and the particle mass \( M=1 \)).
Actually, in liquid \( ^{4}{\textrm{He}} \) the Landau criterion overestimates 
the critical velocity. The explanation of this fact was put forward by Feynman 
(see e.g. \cite{Feynman:smbook}) who suggested that superfluidity is destroyed by 
spontaneous creation of complex excitations (vortex lines, vortex rings, etc.). 
Extensive theoretical and experimental studies on this subject are reviewed in 
\cite{Donelly}. 
 
Bose-Einstein condensation of dilute  trapped clouds of alkali atoms 
\cite{Cor95,Hul95,Ket95} offers new possibilities for the investigation of 
superfluidity \cite{Ketterle:review}. In the spatially homogeneous case, the 
spectrum of elementary excitations of a Bose-condensed gas is given by the 
Bogolyubov dispersion law \cite{Bogolyubov}:
\begin{equation}
\label{Boglaw}
\epsilon (k)=\sqrt{\left(\frac{k^2}{2}\right)^2+2c_{s}^2\frac{k^2}{2}},
\end{equation}
and the Landau critical velocity is equal to the speed of sound
\( c_{s}=\sqrt{gn_{0}} \) (\( n_{0} \) is the condensate density, 
\( g=4\pi a \), and \( a>0 \) is the scattering length). The first experimental
observation of the critical velocity in trapped gaseous condensates, recently
reported by the MIT group \cite{Ketterle:critvel}, gives a significantly
smaller value of \( v_{c} \).
 
The analyses in \cite{Ketterle:critvel} and in recent theoretical publications
(see e.g. \cite{critvel:analit,critvel:numerics} and references therein) employ
the Feynman hypothesis and provide a qualitative explanation of the MIT 
experimental result. In this Letter we point out a simple geometrical effect
which is characteristic for elongated cylindrical traps. Due to the
non-uniform character  of the radial density profile, the spectrum of axially
propagating excitations  in these traps is very different from the Bogolyubov
dispersion law (\ref{Boglaw}), and this difference leads to a strong decrease
of the critical velocity. We show  that this effect can at least partially
explain the small critical velocity  measured in the MIT experiment
\cite{Ketterle:critvel}.   

We consider an infinitely long cylindrical condensate which is harmonically
trapped in the radial (\( \rho  \)) direction. Then the condensate wave function
\( \psi _{0}(\rho ) \) satisfies the Gross-Pitaevskii equation 
\[ 
\left( -\frac{\Delta _{\rho }}{2}+V(\rho
)-\mu +g|\psi _{0}(\rho )|^{2}\right) \psi _{0}(\rho )=0,\]
where \( \mu  \) is the chemical potential, \( V(\rho )=\omega ^{2}\rho ^{2}/2 \)
is the trapping potential, and \( \omega  \) the trap frequency. In the Thomas-Fermi
regime, where the ratio \( \eta =\mu /\omega \gg 1 \), the density profile is given 
by \( n_{0}(\rho )\equiv |\psi _{0}|^{2}=(\mu -V(\rho ))/g \) and the chemical potential 
is related to the maximum condensate density as \( \mu=n_{0max}g \). 
 
Elementary excitations can be regarded as quantized fluctuations of the condensate
wavefunction \cite{LL:volIX}. In our trapping geometry they are characterized by
the axial (\( z \)) wavevector \( k \) and radial angular momentum \( m \).
The corresponding part of the field operator reads  
\[
\delta \hat{\psi }=\sum_{m,k}(u_{mk}\hat{b}_{mk}-v_{mk}^{*}\hat{b}^{\dagger }_{mk}),\]
where \( \hat{b}_{mk}(\hat{b}_{mk}^{\dagger }) \) are annihilation(creation) operators 
of the excitations. The excitation wave functions can be written in the form 
\[
(u,v)_{mk}(\rho ,z)=(u,v)_{mk}(\rho )\exp (im\phi )\exp (ikz),\]
where \( \phi  \) is the angle in the \( x,y \) plane, and the radial functions
\( u_{mk}(\rho ) \) and \( v_{mk}(\rho ) \) are solutions of the Bogolyubov-de
Gennes equations (see e.g. \cite{Ohberg}) 
\begin{eqnarray}     
\epsilon u & = & \left( -\frac{\Delta _{\rho }}{2}+\frac{k^{2}}{2}+V-\mu
+2n_{0}g\right) u+n_{0}gv, \label{BdGu}  \\
-\epsilon v & =& \left( -\frac{\Delta _{\rho }}{2}+\frac{k^{2}}{2}+V-\mu
+2n_{0}g\right) v+n_{0}gu. \label{BdGv}
\end{eqnarray}

Equations (\ref{BdGu}) and (\ref{BdGv}) constitute an eigenvalue problem. For
given \( m \) and \( k \), they lead to a set of frequencies \( \epsilon
_{nm}(k) \) characterized by the radial quantum  number \( n \) which takes
integer values from zero to infinity. In the limit \( \epsilon _{nm}(k)\ll
\mu  \), these modes were found for \( m=0 \) in the hydrodynamic approach in
\cite{Zaremba:soundprop,Kav:soundprop,Stringari:soundprop}.
For \( kR\ll 1 \), where \( R=(2\mu /\omega ^{2})^{1/2} \) is the Thomas-Fermi
radial size of the condensate, the dispersion relation can be expanded in powers
of \( k^{2} \) \cite{Zaremba:soundprop,Stringari:soundprop}: 
\begin{equation}
\label{zarembaexpansion}
\epsilon _{n0}^{2}(k)=2\omega ^{2}n(n+1)+\frac{\omega ^{2}}{4}(kR)^{2}+O(k^{4}).
\end{equation}
The lowest mode (\( n=0 \)) represents axially propagating phonons: For this
mode we have \( \epsilon _{00}(k)=c_{Z}k \), where the sound velocity 
\( c_{Z}=\sqrt{n_{0max}g/2} \) is smaller by a factor of \( \sqrt{2} \) than the 
Bogolyubov speed of sound at maximum condensate density, \( c_{s} \) 
\cite{Zaremba:soundprop,Kav:soundprop,Stringari:soundprop}.
The velocity \( c_{s}/\sqrt{2} \) of axially propagating phonons has been measured
in the MIT experiment \cite{MIT:soundprop}.

The first correction to the linear behavior of the dispersion law \(
\epsilon_{00}(k) \) at \( kR\ll 1 \) reveals its negative curvature: \(
\delta\epsilon_{00}=-\omega (kR)^3/192 \) \cite{Stringari:soundprop}. In
\cite{Zaremba:soundprop} the perturbative analysis leading to
Eq.(\ref{zarembaexpansion}) was extended numerically to \( k\sim 1/R \),
still assuming that \( \epsilon _{n0}(k)\ll \mu  \). The calculations show
that the group velocity of the first mode, \( d\epsilon _{00}/dk \),
monotonously decreases with increasing \( k \) and can become significantly
smaller than \( c_{Z} \) characteristic for \( k\ll 1/R \). This indicates 
that the critical velocity (\ref{Landaucrit}) associated with
creating axial excitations (\( m=0 \), \( n=0 \)) is smaller than \( c_{s} \)
and can also be reduced to below \( c_{Z} \). The physical reason
is that the decrease of the condensate density with increasing \( \rho  \) makes
the axial superfluid flow less stable (see below). 

The hydrodynamic approach used in 
\cite{Zaremba:soundprop,Kav:soundprop,Stringari:soundprop}
is not valid for \( \epsilon \agt \mu  \), where the excitation spectrum is no
longer phonon-like and is dominated by the single particle dispersion relation
\( \epsilon (k)=k^{2}/2 \) (see Eq.(\ref{Boglaw})). The crossover between the
two regimes occurs at \( k\sim k_{c}=\mu ^{1/2} \) and prevents the decrease
of the group velocity with further increase in \( k \). Obviously, the decrease 
of the critical velocity due to the radial inhomogeneity of the density profile 
can be dramatic only for \( k_{c}\gg 1/R \), i.e. in large condensates with 
\( \eta \gg 1 \). 
 
\begin{figure}
{\par\centering \resizebox*{0.9\columnwidth}{!}{\includegraphics{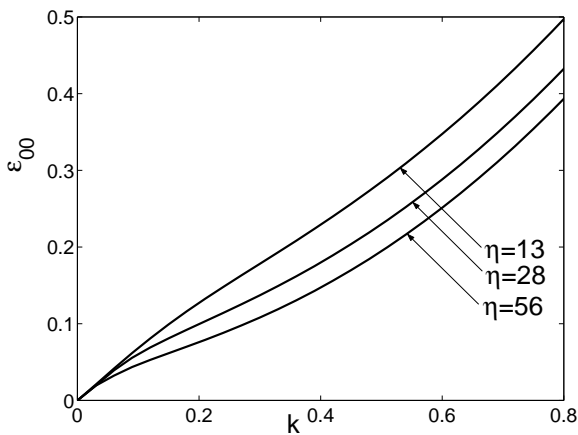}}
\par}
\vspace{4mm} 
\caption{The dispersion law for the lowest mode (\protect\( n=0\protect \), 
\protect\( m=0\protect \)) at various values of \protect\( \eta \protect \).
The excitation energy \protect\( \epsilon_{00}\protect \) is given in units
of \protect\( \mu\protect \), and the momentum \protect\( k\protect \) in units
of \protect\( k_c\protect \). \label{fig:displaw} }
\end{figure}

\begin{figure}
{\par\centering \resizebox*{0.9\columnwidth}{!}{\includegraphics{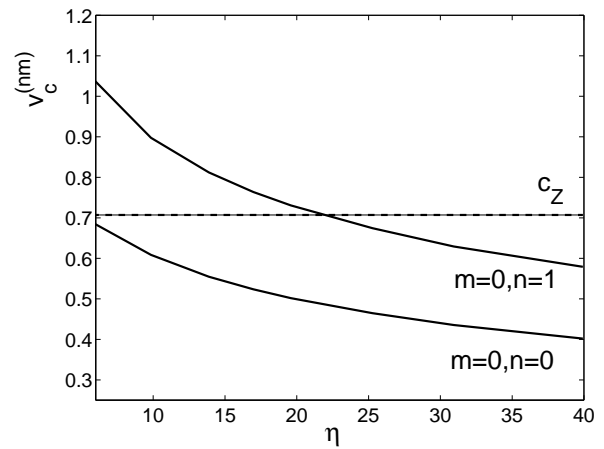}} \par}

\vspace{4mm}

\caption{
The critical velocity 
\protect\( v_c^{(nm)} \protect \) (in units of
\protect\( c_s\protect \)) for the two lowest modes versus the Thomas-Fermi
parameter \protect\( \eta=\mu/\omega \protect \). The dashed line shows the
value of \protect\( c_{Z}\protect \).
\label{fig:crivels} 
} 
\end{figure}

The Bogolyubov-de Gennes equations (\ref{BdGu}) and (\ref{BdGv}) are valid for
arbitrary \( k \) and hence allow us to find the excitation spectrum in the
crossover regime (see Fig.\ref{fig:displaw}) and establish the value of the
critical velocity as a function of the Thomas-Fermi parameter \( \eta  \).
Since the spectrum of excitations consists of a number of independent branches
characterized by the quantum numbers \( n \) and \( m \),
Eq.(\ref{Landaucrit}) gives a value \( v_{c}^{(nm)} \) for the
critical velocity corresponding to each mode. The results of our numerical
calculations for the two lowest modes (\( n=0,\,m=0 \) and \( n=1,\,m=0 \)) are
presented in Fig.\ref{fig:crivels}. The breakdown of superfluidity occurs when 
the velocity of the flow matches the lowest of the velocities \( v_{c}^{(nm)} \), 
which is proved to be \( v_{c}^{00} \). The corresponding curve in 
Fig.\ref{fig:crivels} indicates a significant decrease of the critical velocity 
compared to \( c_{Z} \) already at \( \eta\sim 10 \). 

The decrease of the critical velocity with increasing ratio \( \mu/\omega \) 
seems counterintuitive. One can increase this ratio by decreasing \( \omega \) and 
keeping constant \( \mu \). Then, at a constant density on the axis of the cylinder, 
for radially larger condensates one gets smaller \( v_{c} \). However, this 
phenomenon has a clear physical explanation. The key point is the non-uniform
character of the radial density profile. With increasing axial wavevector \( k \), 
the wave functions of the excitations with \( k\alt k_{c} \) are more localized in the
outer spatial region of the condensate and are thus more sensitive to the
small value of \( n_{0} \) in this region. It is this feature that provides a 
decrease of \( v_{c} \) with increasing \( \eta  \), since for larger \( \eta  \) one 
has more possibilities to increase \( k \) and still satisfy the condition 
\( k\alt k_{c} \). The described situation is not met in liquid helium where the 
density is practically constant and changes only in the region very close to the 
border of the sample. 
 
We now turn to the discussion of the MIT experiment \cite{Ketterle:critvel}.
The radial frequency in this experiment was \( \omega =2\pi \times 65 \) s\(
^{-1} \) and the chemical potential \( \mu =110 \) nK, so that \( \eta \approx 35 \). 
The sound velocity \( c_{s}=6.2 \) mm/s was measured by observing a ballistic expansion.
For these parameters we find \( v_{c}\approx 0.42c_{s}=2.6 \) mm/s, which is not far from 
the observed value \( v_{c}^{(exp)}\approx 1.6 \) mm/s \cite{Ketterle:critvel}.
We do not pretend to explain the MIT data as the condensate in the experiment is not an
infinite cylinder. The ratio of the radial to axial frequency for the MIT trap is \( 3.3 \), 
which indicates that the effect of a finite axial size can be important. Nevertheless, 
our calculations may partially explain the small critical velocity observed in the 
experiment. Moreover, a smaller value of \( v_{c} \) originating from our geometrical effect 
can facilitate the nucleation of vortex objects and thus provide a larger decrease of the
critical velocity.

We would like to thank M. Lewenstein, A. Sanpera and A. Muryshev for fruitful
discussions and numerical insights. The work was supported by the Austrian Science
Foundation, by the Stichting voor Fundamenteel Onderzoek der Materie (FOM),
by INTAS, and by the Russian Foundation for Basic Studies (grant 99-02-1802).

\bibliographystyle{prsty}
\bibliography{myDb}

\end{document}